\documentclass[a4paper,twoside]{article}

\usepackage{epsfig}
\usepackage{subcaption}
\usepackage{calc}
\usepackage{amssymb}
\usepackage{amstext}
\usepackage{amsmath}
\usepackage{amsthm}
\usepackage{multicol}
\usepackage{pslatex}
\usepackage{apalike}
\usepackage[ruled,linesnumbered]{algorithm2e}
\usepackage{algcompatible}
\usepackage{algpseudocode}
\algdef{SE}{Begin}{End}{\textbf{Begin}}{\textbf{End}}
\algnewcommand{\TRUE}{\textbf{true}}
\algnewcommand{\False}{\textbf{false}}
\algnewcommand{\Parameters}{\textbf{Parameters}}
\usepackage[bottom]{footmisc}
\usepackage{SCITEPRESS}     

\begin{document}

\title{IRatePL2C: Importance Rating-based Approach for Product Lines Collaborative Configuration}

\author{\authorname{Sihem Ben Sassi\sup{1}\orcidAuthor{0000-0002-1925-4989}}
\affiliation{\sup{1}RIADI Lab., National School of Computer Science, Manouba University, Tunisia\footnote{preprint - accepted in ENASE2024 \\Ben Sassi, S. (2024). IRatePL2C: Importance Rating-Based Approach for Product Lines Collaborative Configuration.  In Proceedings of the 19th International Conference on Evaluation of Novel Approaches to Software Engineering, ISBN 978-989-758-696-5, ISSN 2184-4895, pages 784-791.}}
\email{sihem.bensassi@gmail.com}
}


\abstract{Collaborative configuration of product lines has appealed the interest of several researchers. 
Some of them proposed an approach in which involved stakeholders can freely configure the product line without being constrained by the choices made the other ones. The core of any proposed approach in this context focuses on how conflictual situations are resolved. Few works consider stakeholders preferences in their resolution process. However, to generate a valid solution satisfying all constraints, they generally rely on a process of exponential complexity. In this work, we propose the IRatePL2C approach, which resolution strategy relies on importance degrees assigned by the stakeholders to their initial configuration choices. IRatePL2C starts by merging stakeholders' configurations and then detecting and resolving the conflicts according to their type: explicit or implicit in sequential steps. Finally, domain constraints are propagated and the process is reiterated to reach a final valid configuration. An illustrative example is presented to evaluate the approach. The complexity of IRatePL2C is polynomial which an important advantage compared with previous works.
}

\onecolumn \maketitle \normalsize \setcounter{footnote}{0} \vfill

\section{\uppercase{Introduction}} \label{sec:introduction}
Collaborative configuration of product lines refers to a coordinated process in which a set of stakeholders (e.g. product manager, domain expert, software engineer) share the configuration activities based on their areas of expertise in order to decide about features that should be included in the final product \cite{jss2019}. It has appealed the interest of several researchers, such as \cite{ref7}, \cite{ref6}, \cite{ref4}, \cite{ref9}, \cite{ref8}, \cite{ref10}, \cite{stein2014}, \cite{ref11}, \cite{ref13}, \cite{enase2020}, \cite{viet2022} and \cite{jss2023}. 
The mainspring of the proposed approaches is to handle situations where several stakeholders are involved in the configuration of a product line, and therefore to manage conflictual situations that may arise when these stakeholders have contradictory choices regarding the final product. 

The prevailing means to represent product lines is feature model \cite{kevin}, \cite{terstudy}, in which features are related with mandatory, optional, alternative (XOR) and OR relationships. Furthermore, the feature model is accompanied with inclusion (require) and exclusion (exclude) relationships capturing domain constraints \cite{Benavides2010}. In this case, product configuration consists in choosing (selecting) a set of features that meet individual requirements toward the desired product. When the configuration is made by a single stakeholder, it is easy to ensure its validity, i.e. conformance to the product line model with its domain constraints \cite{ref1}, thanks to constraints propagation \cite{czarnecki2005}. However, when several stakeholders are involved in this process, their requirements may be contradictory or do not comply with the product model; such situations of inconsistency are referred to as conflicts \cite{osman2009}. Furthermore, in collaborative context, stakeholders requirements are better captured by allowing them to choose the undesired features ({$\neg F_j$}) that should not be present in the final product as well as the desired ones ({$F_i$}) \cite{stein2014}, \cite{viet2022}, \cite{jss2023}. 

According to \cite{enase2020}, a conflict may be (i) explicit when  the same feature is explicitly desired by a stakeholder and undesired by another; (ii) implicit when configuration choices of the stakeholders violate the product line feature model constraints. Two cases are distinguishable: (ii-1) two alternative features (related with XOR) are desired by different stakeholders; and (ii-2) domain constraint propagation makes the configuration not valid because it requires either to include a feature that is already undesired by another stakeholder or to exclude a feature that is desired by another stakeholder. 

Obtaining a valid configuration from the different stakeholders' configuration choices requires to resolve any detected conflict by eliminating one or more configuration choices involved in the conflict. Existing approaches tackle this issue by (1) supporting either workflow-based process imposing a predefined configuration order, or a flexible one allowing a free order configuration, where stakeholders either configure the whole product line model, or just an assigned part of it; and (2) implementing a conflict resolution method adopted from social science like negotiation and predefined priority, or more technical one such as range fixes. A comprehensive view characterizing collaborative configuration approaches with respect to a comparison framework is given in \cite{jss2019}. Actually, few works propose a flexible configuration process that takes into account stakeholders preferences in conflict resolution, namely \cite{stein2014}, \cite{Ochoa15}, \cite{viet2022} and \cite{jss2023}. However, either they do not generate a solution that takes into account the preferences of all stakeholders, and / or they implement an expensive computation process. 

The remainder of this paper is structured as follows: Section 2 is dedicated to the proposed approach and the details of its process. Section 3 illustrates IRatePL2C with a full example. Section 4 discusses the stakeholders satisfaction and the complexity of the solution compared to related works. Finally, Section 5 concludes the paper.

\section{\uppercase{The IRatePL2C Approach}}
The proposed IRatePL2C approach is based on the following principles and hypotheses:
 
 \begin{itemize}
	\item \textbf{H1}: each stakeholder freely configures the product line; there is no predefined order between stakeholders during the configuration process;
\item \textbf{H2}: each stakeholder makes a configuration decision by specifying the un/desired features features to be absent/present in the product;
	\item \textbf{H3}: a prior agreement between stakeholders to express to which extent they desire a configuration choice i.e.  a (un)desired feature (not) to be included in the final configuration.
	\item \textbf{H4}: each stakeholder assigns an importance degree to each of his/her explicit configuration choices; 
	 \item \textbf{H5}: a five scale-based score is used to express the configuration choice importance degree, where:
		5 means that the choice is very important to be considered in the final configuration; 4: fairly important; 3: important; 2: slightly important; and 1: not at all important.
	\item \textbf{H6}: importance degrees are only assigned to the explicit choices (initial choices) made by the stakeholder. In other words, features that should be included in the configuration because of the constraints propagation to ensure its validity do not receive scores.
	   \item \textbf{H7}: each stakeholder is aware that potential conflicting choices will be resolved based on the importance degrees, and assigns consciously scores to the configuration choices he/she makes. 
 \end{itemize}
 
\begin{figure*}[!h]
  \centering
   \includegraphics [width=15cm]{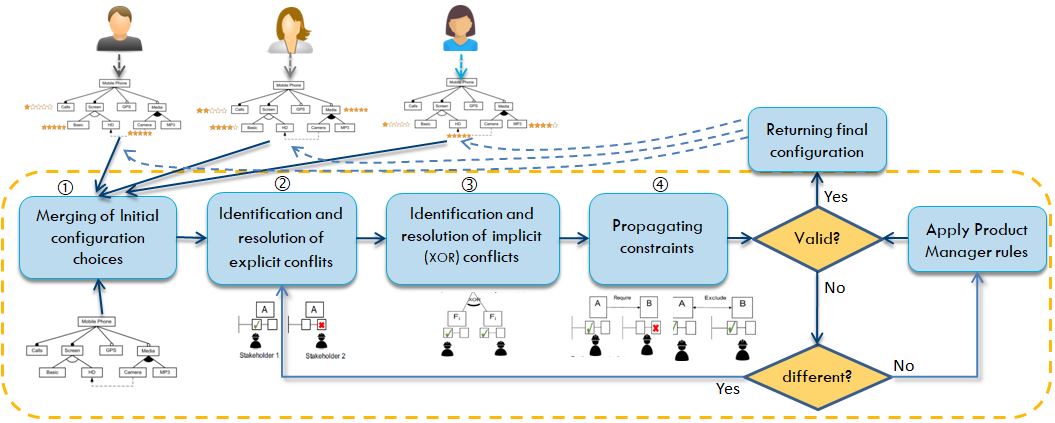}
  \caption{The proposed approach process.}
  \label{FigApproach}
 \end{figure*}

Figure \ref{FigApproach} shows the main steps of IRatePL2C approach: it starts by (1) merging the initial configuration choices made by each stakeholder; as pointed earlier, stakeholders assign an importance degree to each explicit choice they made. Second (2), the merged configuration is analyzed to identify explicit conflicts; these latter are resolved and the configuration is accordingly updated. Third (3), the new configuration is analyzed to identify and resolve implicit conflicts resulted from XOR constraints; the configuration is updated accordingly. Fourth (4), the product line domain constraints are checked to propagate them on the configuration when applicable. The configuration is afterward checked. If it is valid, it will be returned as final configuration. When the configuration is not valid, it is question to compare it with the one of the previous iteration in order to decide to reiterate the process starting from Step -2- in the case the two configurations are different, or hand over to the product line manager in order to apply the rule he/she advocates to resolve remaining conflicts. This rule may be for example one of the substitution rules proposed in \cite{enase2020} and \cite{jss2023}, that resolves conflicts, among others, by favoring "the most complete product``, or "the simplest product``, or else by prioritizing "the explicit choices made by a given stakeholder``.

Algorithm \ref{algorithm1} summarizes the core process of IRatePL2C approach. It requires the feature model of the product line to configure as a tree (\textit{PL\_FM}), the set of constraints related to that model including XOR, include and require ones (\textit{PL\_ConstraintsList}), as well as the list of stakeholders involved in the collaborative configuration (\textit{Stk\_List}). As result, it ensures a valid configuration as a list of features deduced from stakeholders' configuration choices (\textit{FinalConfig}). Statements 2-5 initialize four variables: two Boolean ones $valid$ and $different$ to store the result of checking respectively the validity of the configuration and the alteration of the configuration from one iteration to another; and two lists to store respectively configuration choices to remove from the configuration in order to resolve a set of conflicts $toRemoveList$, and the set of non resolved conflicts $RemainedConflictsList$. 
Statements 6-8 allow to collect stakeholders' configurations. 
IRatePL2C approach's step -1- is reflected by Statement 9 which invokes $MergeConfigs()$ procedure. The $while$ loop encompasses the three next  steps implemented thanks to the $ResolveExplicitConflicts()$ procedure (Statement 13) for Step -2-, $ResolveXORConflicts()$ procedure (Statement 16) for Step -3-, and the $PropagateConstraints()$ function (Statement 19) for Step -4-. The two procedures have as output a list of configuration choices to remove from the current configuration $CurrentConfig$ in order to resolve the related identified conflicts (Statements 14 and 17). The function determines the (un)desired features to add the current configuration (Statement 20) and updates the configuration choices importance dgrees accordingly (Statement 21). The $while$ loop exits either with a valid configuration, in such case it represents the $FinalConfig$ to return, or with a configuration that still has some remaining conflicts where importance degrees cannot help in their resolution; in such case, the product manager takes a decision and final configuration is returned to stakeholders. 

\begin{algorithm*}[!h]
 \caption{Core Process of IRatePL2C Approach.}
\label{algorithm1}
 \KwData{
\textit{PL\_FM}: the tree of features representing the product line to configure\\
         \textit{PL\_ConstraintsList}: the list of constraints related to the product line\\
				 \textit{Stk\_List}: the list of stakeholders involved in the configuration}
 \KwResult{
\textit{FinalConfig}: a valid configuration as a list of features deduced from stakeholders' choices }
\Begin {
  $valid \leftarrow False $ \;
  $different \leftarrow True $ \;
	$toRemoveList \leftarrow \emptyset $ \;
	$RemainedConflictsList \leftarrow \emptyset $ \;
	
\ForEach { $ (Stk_{i} \in Stk\_List)$}{
		$LE\_ConfigStk_{i} \leftarrow Configure(PL\_FM) $ \;
	}
 
$ MergeConfigs(\{LE\_ConfigStk_{i} \} ~i=1..Nb\_Stks \rightarrow List\_EFs, List\_EFsImportance)$ \;

 $CurrentConfig \leftarrow List\_EFs $ \;

 \While{(not valid and different)}{
    $PreviousConfig \leftarrow CurrentConfig$\;
		$ResolveExplicitConflicts(CurrentConfig, List\_EFsImportance \rightarrow toRemoveList, RemainedExplicitConflictsList) $\;
		$CurrentConfig \leftarrow RemoveFromList(CurrentConfig, toRemoveList) $\;
		$RemainedConflictsList \leftarrow RemainedConflictsList \cup  RemainedExplicitConflictsList $ \;
		$ResolveXORConflicts(CurrentConfig, List\_EFsImportance, PLConstraintsList \rightarrow toRemoveList, RemainedXORConflictsList)$\;
		$CurrentConfig \leftarrow RemoveFromList(CurrentConfig, toRemoveList) $\;
		$RemainedConflictsList \leftarrow RemainedConflictsList \cup  RemainedXORConflictsList $ \;
		
		$toAddList \leftarrow PropagateConstraints(CurrentConfig, PL\_ConstraintsList)$\;
		$CurrentConfig \leftarrow AddToList(CurrentConfig, toAddList)$\;
		$List\_EFsImportance \leftarrow UpdateFImportance(List\_EFsImportance, toAddList)$\;
		$different \leftarrow CompareConfigLists(PreviousConfig, CurrentConfig)$\;
		$valid \leftarrow CheckValidity(CurrentConfig)$\;
		}
	\eIf{valid}{
     $FinalConfig \leftarrow CurrentConfig $\;
   } (\tcc*[f]{not valid and not different}) 
	{
    $RemainedConflictsList \leftarrow CheckRemainedConflicts(CurrentConfig, RemainedConflictsList)$ \; 
		$FinalConfig \leftarrow ApplyProductManagerRules(CurrentConfig, RemainedConflictsList)$\;
   }
   \Return{FinalConfig};
 }
\end{algorithm*}

\subsection{Configurations Merging}
\begin{figure*}[!h]
  \centering
   \includegraphics [width=8cm]{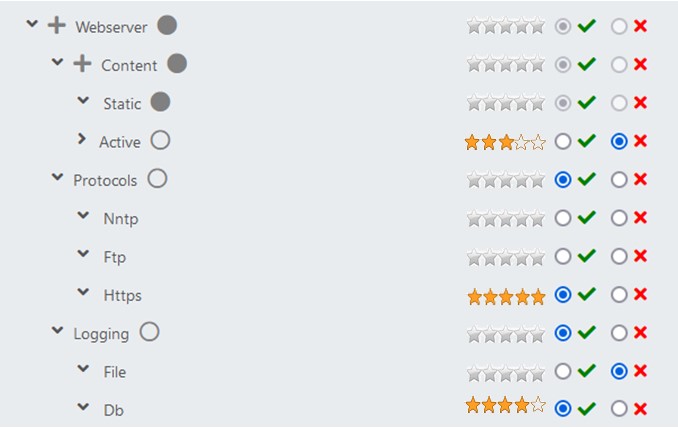}
  \caption{Extract of a stakeholder's configuration choices.}
  \label{FigRateEx}
 \end{figure*}

As mentioned earlier, each stakeholder makes a set of explicit configuration choices expressing the features he/she desires to be present or not to be present in the final configuration. Each explicit choice is accompanied by an importance degree ranging from 1 to 5. Only explicit choices made by the stakeholder are subject to importance degree assignment. Figure \ref{FigRateEx} shows an extract of a configuration where the stakeholder explicitly selected $\neg Active$ (i.e. does not desire this feature to be present in the final configuration) with a choice importance equals to 3; he/she also selected $https$ and $Db$ with choice importance equal to 5 and 4 respectively. Figure \ref{FigRateEx} shows also that even though $\neg File$ is selected, no importance assignment is considered. Indeed, this choice was not explicitly made by the stakeholder, but generated thanks to constraints propagation: the $alternative$ relationship between the two features $File$ and $Db$ in this case since $Db$ was explicitly selected.
 
Algorithm \ref{AlgoMerge} describes the process of configuration merging, which consists in obtaining, from the stakeholders configuration choices $LE\_ConfigStk_{i}$, two lists: (1) the set of all explicit configuration choices including the desired features {$F_i$} and non desired features {$\neg F_j$} without redundancy $List\_EFs$ (Statement 7), and (2) the set of all explicit configuration choices, where each one of them is accompanied with an ordered list of its related importance degrees expressed by stakeholders who made that choice $List\_EFsImportance$ (Statements 8 and 10).
 
\begin{algorithm*}[!h]
 \caption{MergeConfigs: Stakeholders initial Configurations Merging.}
\label{AlgoMerge}
 \KwData{
\textit{$ LE\_ConfigStk_{i} ~i=1..Nb\_Stks $}: list of initial configuration choices made by each stakeholder}
 \KwResult{
\textit{List\_EFs}: a list of features representing the set of merged configuration choices of the stakeholders\\
 \textit{List\_EFsImportance}: a list of the merged configuration choices along with the importance degrees declared by the stakeholders }

\Begin {
$ List\_EFs \leftarrow \emptyset $ \;
  $ List\_EFsImportance \leftarrow \emptyset $ \;
	\ForEach { $ LE\_ConfigStk_{i} ~i=1..Nb\_Stks $}{
   \ForEach { $ e \in LE\_ConfigStk_{i} $}{
      \eIf{not (e.F  $\in$ List\_EFs)}{
          $List\_EFs \leftarrow List\_EFs \cup e.F $\;
					$List\_EFsImportance \leftarrow List\_EFsImportances \cup (e.F, e.I) $\;
         }
			{ $ List\_EFs \leftarrow insertOrderedImportance(List\_EFs, e.F, e.I) $\;
		
			}
		}
   }
 }
\end{algorithm*}

\subsection{Explicit Conflicts Resolution}
This step consists in detecting explicit conflicts in the merged configuration choices resulted from the previous step, then resolving them using importance degrees comparison of involved features. As mentioned in the introduction, an explicit conflict occurs when the same feature is, at the same time, desired by a stakeholder to be part of the configuration and undesired by another stakeholder.  

The resolution is done by considering importance degrees for the configuration choices expressed by stakeholders. A configuration choice corresponds to the feature state, i.e. desired ($f$), undesired ($\neg f$). If the configuration choice is present only one time (meaning that only one stakeholder desires to include or exclude that feature), the same importance degree expressed by that stakeholder is reported. In case that several stakeholders desired the same state regarding a given feature, the corresponding importance degree is based on the ordered importance given by the stakeholders related to that feature. The rationale is that we need to keep what each stakeholder believes essential to include in the configuration. We resolve conflicts by keeping the configuration choice with the highest importance degree, and eliminating from the configuration the other one involved in the conflict.

Algorithm \ref{algoExplicitConf} describes the explicit conflicts resolution steps. It requires as input a configuration in an intermediate state $CurrentConfig$ and the list of the merged configuration choices with their related importance degrees $List\_EFsImportance$. It returns two lists; the first one is the list of configuration choices to delete from the configuration $CurrentConfig$ to resolve the identified explicit conflicts ($toRemoveList$); the second one contains any eventual non resolved explicit conflict ($RemainedExplicitConflictsList$). 
Statement 10 compares the importance degrees of a configuration choice $f$ with those of its negation $\neg f$ extracted from $List\_EFsImportance$ in Statements 8 and 9. The comparison consists in walking through the two lists degrees; since they are ordered, the first who has an element of grater value than the corresponding one in the other list, or who has a greater number of elements in case of equality is the winner. The looser is added to $toRemoveList$ (Statements 11-15). In case of equality in values and lists length, the conflict is added to $RemainedConflictsList$ (Statement 17). This process is applied to each configuration choice in the form of $\neg f$ in the $CurrentConfig$ in order to detect all conflicts (Statements 4-5).
The current configuration is afterward updated by removing $toRemoveList$ from $CurrentConfig$ (Statement 14 in Algorithm \ref{algorithm1}).

\begin{algorithm*}[!h]
 \caption{ResolveExplicitConflicts: Determine explicit conflicts and which feature state to remove.}
\label{algoExplicitConf}
 \KwData{
\textit{CurrentConfig}: a list of features reflecting an intermediate state of a configuration \\
\textit{List\_EFsImportance}: a list of the merged configuration choices along with the importance degrees declared by the stakeholders}
 \KwResult{
\textit{toRemoveList}: a list of features to remove from $CurrentConfig$ to resolve detected explicit conflicts\\
 \textit{RemainedExplicitConflictsList}: a list of explicit conflicts that has not been resolved
}

\Begin {
$ toRemoveList \leftarrow \emptyset $ \;
  $ RemainedExplicitConflictsList \leftarrow \emptyset $ \;
	
   \ForEach { $ nf \in CurrentConfig $}{
      \If{nf.charAt[0]='$\neg $')}{
          $f \leftarrow nf.subString(1) $\;
					\If{($f \in CurrentConfig$)}{
					     $f\_ImpList \leftarrow extract(List\_EFsImportances, f) $\;
							 $nf\_ImpList \leftarrow extract(List\_EFsImportances, nf) $\;
							 $res \leftarrow isMoreImportant(f\_ImpList, nf\_ImpList) $\;
							 \eIf{(v=1)} {
							    $toRemoveList \leftarrow toRemoveList \cup nf $\;}
									 {\eIf {(v=2)} {
							          $toRemoveList \leftarrow toRemoveList \cup f $\;}
												{$ RemainedExplicitConflictsList \leftarrow RemainedExplicitConflictsList \cup (f, nf) $\;}
								    }
					}
	    	}
   }
 }
\end{algorithm*}

\subsection{Implicit Conflicts Resolution}
The next step in the resolution focuses on implicit conflicts, especially when alternative features (related
with XOR) are desired by different stakeholders. The resolution principle here is similar to the used one for the explicit conflicts, but applied to features involved in violated XOR constraints. As it is illustrated in Statements 4-7 of Algorithm \ref{AlgoXor}, it is first question to identify any two features participating in a XOR constraint. Then, extract the importance degrees associated to each one (Statements 9-10) to determine which one to remove (Statements 11-16), or add the conflict to $RemainedXORConflictsList$ if necessary. The configuration choices in the returned $toRemoveList$ are removed from the configuration $CurrentConfig$  and the non resolved conflicts are added to $RemainedConflictsList$ (Statements 17 and 18 in Algorithm \ref{algorithm1}).

\begin{algorithm*}[!h]
 \caption{ResolveXORConflicts: Determine XOR conflicts and which feature to remove.}
\label{AlgoXor}
 \KwData{
\textit{CurrentConfig}: a list of features reflecting an intermediate state of a configuration \\
\textit{List\_EFsImportance}: a list of the merged configuration choices along with the importance declared by the stakeholders \\
\textit{PLConstraintsList}: the list of constraints related to the product line}
 \KwResult{
\textit{toRemoveList}: a list of features to remove from $CurrentConfig$ to resolve detected explicit conflicts\\
 \textit{RemainedXORConflictsList}: a list of implicit XOR conflicts that has not been resolved 
}

\Begin {
$ toRemoveList \leftarrow \emptyset $ \;
  $ RemainedXORConflictsList \leftarrow \emptyset $ \;
	
   \ForEach { $ Constraint_{i} \in PLConstraintsList $}{
      \If{(TypeOf($Constraint_{i}$)=XOR )}{
			    \For{$j\leftarrow 1$ \KwTo nbFeatures($Constraint_{i}$)-1} {
					    \For{$m\leftarrow j+1$ \KwTo nbFeatures($Constraint_{i}$)} {
					   				\If{(($F_{j} \in CurrentConfig$) and ($F_{m} \in CurrentConfig$))}{
												$fj\_ImpList \leftarrow extract(List\_EFsImportances, F_{j}) $\;
												$fm\_ImpList \leftarrow extract(List\_EFsImportances, F_{m}) $\;
												$v \leftarrow isMoreImportant(fj\_ImpList, fm\_ImpList) $\;
											\eIf{(v=1)} {
													$toRemoveList \leftarrow toRemoveList \cup F_{m} $\;}
													{\eIf {(v=2)} {
															$toRemoveList \leftarrow toRemoveList \cup F_{j} $\;}
															{$ RemainedXORConflictsList \leftarrow RemainedXORConflictsList \cup (F_{m}, F_{j}) $\;}
								           }
					          }
	            	}
           }
				}
		}
 }
\end{algorithm*}

\subsection{Constraints propagation}
For the someone to ask why the next step does not deal with resolving the second type of implicit conflicts resulting from \textit{require} or \textit{exclude} constraints violation. Actually, a \textit{require} constraint (say $f_i \Rightarrow f_j$)  implies the inclusion of the feature of its second part ($f_j$) when the one of its first part ($f_i$) is present in the configuration (i.e. desired by a stakeholder). The conflict occurs when the second feature to be included ($f_j$) is already undesired by another stakeholder ( i.e. $ \neg f_j$ is part of the configuration). This means that the configuration encompasses a feature and its negation ($f_j, \neg f_j$) which amounts to resolving explicit conflicts.

Similarly, an \textit{exclude} constraint (say $f_i \Rightarrow \neg f_j$)  implies a mutual exclusion between the two features: if $f_i$ is present in the configuration, $f_j$ should not, and hence $\neg f_j$ should be included; if $f_j$ is present in the configuration, $f_i$ should not, and hence $\neg f_i$ should be included. A conflict occurs when $f_i$ (resp. $f_j$) is selected by one stakeholder; $\neg f_j$ (resp. $\neg f_i$) is therefore included in the configuration while $f_j$ (resp. $f_i$) is already present as desired by another stakeholder. This also means that the configuration encompasses a feature and its negation ($f_j, \neg f_j$) (resp. ($f_i, \neg f_i$)) which amounts as well to resolving explicit conflicts.

Constraints propagation is already dealt with in the literature \cite{czarnecki2005}, \cite{Benavides2010}; it is simply done by walking through domain (\textit{require} and \textit{exclude}) constraints and updating the configuration by adding a feature or its negation accordingly. For the purpose of this work, we need to consider the importance degree of any desired or undesired feature added to the configuration in order to keep being able to resolve conflicts based on importance degrees. As the source of any addition is a configuration choice belonging to the configuration and causing the \textit{require} or \textit{exclude} constraint to be triggered, we assign to the added (un)desired feature the highest importance degree (MAX) associated with the source configuration choice. 
Statement 19 in Algorithm \ref{algorithm1} returns the list of (un)desired features to add to the current configuration (Statement 20). This same list is also used to update the importance degrees associated with the  configuration choices already made by stakeholders and \/ or add non existing (un)desired features with their related importance degrees (Statement 21).

On the other hand, any newly added feature to the configuration may violate a XOR constraint. This explains the reason behind the iterative processing for the conflicts resolution in IRatePL2C approach: looping on resolving explicit conflicts, resolving implicit XOR conflicts and propagating (\textit{require} and \textit{exclude}) constraints. 

Furthermore, it is obvious that the resolution of a given conflict may result in the resolution of another conflict. For example, the resolution of an explicit conflict may also resolve a conflict caused by a XOR constraint. This also explains the rationale behind the stepped resolution of conflicts. In each step, conflicts of a certain type are detected and resolved before dealing with another type. This also explains why remained non resolved conflicts need to be checked before passing them to the product manager in Statements 26-27 of Algorithm \ref{algorithm1}.

\section{\uppercase{Illustrative Example}}
\begin{figure*}[!h]
  \centering
   \includegraphics [width=15cm]{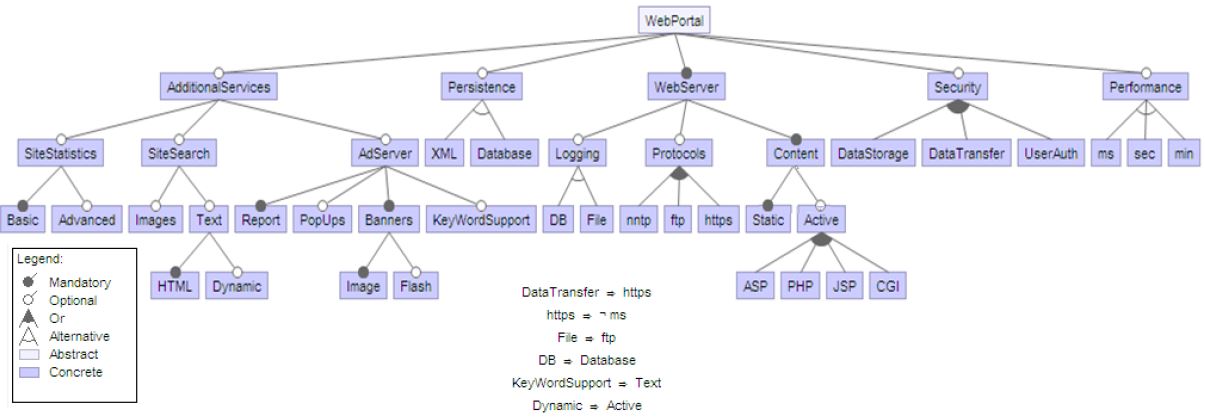}
  \caption{Web Portal Feature Model.}
  \label{FigWPortal}
 \end{figure*}

In this section, we illustrate the IRatePL2C approach through an example of a product line feature model used in previous works dealing with collaborative configuration (e.g. \cite{ref4} and \cite{jss2023}). It is question of the Web portal feature model shown in Figure \ref{FigWPortal}. The model contains three XOR relationships between $XML$ and $Database$, between $DB$ and $File$ and between $ms$, $sec$ and $min$. It comes also with set of five domain constraints of inclusion requirement (e.g. $DataTransfer$ requires $https$; $Dynamic$ requires $Active$); and one exclusion requirement (namely, $https$ excludes $ms$). 

In the example scenario, we assume that five stakeholders collaboratively configure the Web portal model. Each of them freely express his/her configuration choices as desired and undesired features. With each explicit choice made, the stakeholder assigns a related importance degree. Table \ref{exampleScenario} summarizes the configuration scenario; the importance degree is noted as superscript of the related configuration choice. 

\begin{table} [htb!]
 \centering
        \vspace{-0.2cm}
        \caption{Collaborative configuration scenario for the Web Portal model.}  \centering
				\scriptsize{
       \begin{tabular}{|c|p{5cm}|}
   \hline Stakeholder & Explicit configuration choices  \\
   
      \hline 
      Stk1 & KeyWordSupport$^{(2)}$, DB$^{(4)}$, $\neg$Active$^{(3)}$, https$^{(5)}$  \\ \hline
      Stk2 &  XML$^{(4)}$, $\neg$Text$^{(4)}$, $\neg$Active$^{(5)}$,  ms$^{(3)}$\\ \hline
      Stk3 & Active$^{(5)}$, Php$^{(2)}$, XML$^{(1)}$, DataTransfer$^{(4)}$ \\ \hline
	    Stk4 & Text$^{(2)}$, Dynamic$^{(5)}$, KeyWordSupport$^{(4)}$, DB$^{(3)}$,  $\neg$https$^{(1)}$, $\neg$Sec$^{(3)}$ \\ \hline
	    Stk5 & Text$^{(4)}$, Database$^{(5)}$, Active$^{(4)}$, DataTransfer$^{(3)}$ \\  \hline

\end{tabular}
}
\label{exampleScenario}
\end{table}

It is worth mentioning that throughout this section we omit, for the sake of clarity and space, to include in the configurations the mandatory features, namely WebServer, Content and Static. Furthermore, we mainly focus on features susceptible to cause conflicts; the other ones are also omitted for the same reason. For example, to have $XML$ in the configuration, its parent feature $Persistence$  should also be part of the configuration; to have $Text$ in the configuration, both $AdditionalServices$ and $SiteSearch$ must also be part of it as well as the mandatory $HTML$ feature.

\textbf{Step-1: merging stakeholders' configuration choices}\\
As described in the previous section, merging stakeholders' configuration choices leads to two results. The first one is the set of the configuration choices without redundancy ($List\_EFs$) and the second one associates to each configuration choice the ranked list of its related importance degrees ($List\_EFsImportance$). These two results are shown in Table \ref{FeatureImportance}. For example, both Stk1 and Stk4 desire $KeyWordSupport$ with an importance degree equals to 2 and 4 respectively. The list (4, 2) is therefore associated as ordered importance degrees to the configuration choice $KeyWordSupport$.

\begin{table} [htb!]
 \centering
        \caption{Merged configuration choices with their importance degrees results.}  \centering
				\scriptsize{
       \begin{tabular}{|p{2.25cm}|p{1cm}||p{1.5cm}|p{1cm}|}
			\hline\multicolumn{4}{|c|}{Merged configuration choices}\\
			\hline\multicolumn{4}{|p{5.75cm}|}{KeyWordSupport, DB, https, XML, ms, Active, Php, $\neg$https, DataTransfer, Text, Dynamic, Database, $\neg$Text, $\neg$Active,  $\neg$Sec} \\ \hline 
			\hline\multicolumn{4}{|c|}{Configuration choices with importance degrees}\\ \hline 
   \hline Feature & degree & Feature & degree\\    
      \hline 
https	&5 & $\neg$https&	1 \\ \hline
Active	&5, 4 & $\neg$Active	&5, 3 \\ \hline
Text&	4, 2& $\neg$Text	&4 \\ \hline
ms	&3 & $\neg$Sec	&3 \\ \hline
XML&	4, 1 & PHP&	2 \\ \hline
DataTransfer	&4, 3&  Dynamic	&5 \\ \hline
KeyWordSupport	&4, 2 & Database	&5 \\ \hline
DB	&  4, 3 & &  \\ \hline
\end{tabular}
}
\label{FeatureImportance}
\end{table} 

\textbf{Step-2: resolving explicit conflicts}\\
Explicit conflicts resolution starts by detecting detecting explicit conflicts. In this example, three conflicts of this type are identified, as described in Table \ref{ExplicitConfR}. To decide which feature state to retain and which one to remove, we use the importance degrees ordered lists associated with each configuration choice. In the case of the first conflict, $https$ has one importance degree which is 5, while $\neg https$ has 1 as sole importance degree; this means that $https$ is retained and $\neg https$ has to be removed. As to $Active$ and $\neg Active$, both of them have 5 as first importance degree; the following degree should therefore be considered. The second importance degree is respectively 4 and 3, which means to keep $Active$ in the configuration and remove $\neg Active$. Regarding $Text$ and $\neg Text$, both have 4 as first importance degree; however $\neg Text$ has only one importance degree while $Text$ has two ones. This means that $Text$  is retained and $\neg Text$ should be removed. The configuration is updated accordingly as reported in Table \ref{ExplicitConfR}.

\begin{table} [htb!]
 \centering
        \caption{Explicit conflicts resolution results.}  \centering
				\scriptsize{
       \begin{tabular}{|p{2cm}|p{2cm}|p{2cm}|}
			
   \hline Detected conflict & Retained choice & To Remove choice\\    
      \hline 
(https, $\neg$https) &	https &	$\neg$https \\ \hline
(Active, $\neg$Active)	& Active  & $\neg$Active\\ \hline
(Text, $\neg$Text)	&Text  & $\neg$Text\\ \hline
\hline\multicolumn{3}{|c|}{Updated configuration}\\
\hline\multicolumn{3}{|p{5.6cm}|}{KeyWordSupport, DB, https, XML, ms, Active, Php, DataTransfer, Text, Dynamic, Database, $\neg$Sec} \\ \hline
\end{tabular}
}
\label{ExplicitConfR}
\end{table} 

\textbf{Step-3: resolving implicit XOR conflicts}\\
XOR constraints are in this step checked to detect if they are violated. In this example, one of the three XOR constraints is not respected. Indeed, both $XML$ and $Database$ are included in the configuration while they are related with XOR. Importance degrees determined in Step-1 and shown in Table \ref{FeatureImportance} are used to decide which feature to remove. The importance degrees associated with $XML$ are 4 and 1. Even though $Database$ has only one importance degree, but it wins as is importance degree which is 5 is greater than 4. $XML$ should therefore be eliminated from the configuration.
The resulted configuration is therefore:
\noindent
\begin{align*}
&(KeyWordSupport, DB, https,   ms, Php, Text,\\
 &Dynamic, \neg Sec, Database, Active, DataTransfer)
\end{align*}


\textbf{Step-4: propagating constraints}\\
Based on the configuration resulted from the previous step and on the $require$ \/ $exclude$ constraints associated with the Web portal feature model shown in Figure \ref{FigWPortal} as well as on configuration choices importance degrees depicted in Table \ref{FeatureImportance}, we easily see for example that $KeyWordSupport$ causes to add $Text$ with importance degree equals to 4 and $https$ causes to add $\neg ms$ with 5 as importance degree. 
The full list of added (un)desired features are clarified in Table \ref{ConstProp}. The importance degrees associated with the configuration choices and the newly (un)desired added features in updated accordingly.
The new resulted configuration is as follows:
\noindent
\begin{align*}
&(KeyWordSupport, DB, https, \neg ms,  ms, Php, Text,\\
 &Dynamic, \neg Sec, Database, Active, DataTransfer)
\end{align*}

\begin{table} [htb!]
 \centering
        \caption{Constraints propagation results.}  \centering
				\scriptsize{
       \begin{tabular}{|p{3cm}|p{1.5cm}|p{1.5cm}|}
         \hline 
Constraint &	(Un)desired feature &	Importance degree \\ \hline
	(KeyWordSupport $\Rightarrow$ Text) & Text & 4 \\ \hline
  (DB $\Rightarrow$ Database) & Database & 4 \\ \hline
	(https $\Rightarrow$ $\neg$ ms) & $\neg$ ms & 5 \\ \hline
	(Dynamic $\Rightarrow$ Active) & Active & 5 \\ \hline
	(DataTransfer $\Rightarrow$ https) & https  & 4 \\ \hline
\end{tabular}
}
\label{ConstProp}
\end{table}

The process needs to carry on as the last obtained configuration is not valid yet and is different from the first one. In the second iteration, (i) explicit conflicts resolution reveals that the conflict ($ms, \neg ms$) is resolved by keeping $\neg ms$ since the latter has 5 as importance degree versus 3 for $ms$; (ii) all XOR constraints are respected; (iii) constraints propagation does not alter the configuration obtained from the previous step; (iv) the validity is checked and verified. The process stops and returns the final configuration based on the following configuration decisions.
\begin{align*}
&(KeyWordSupport, DB, https, \neg ms, Php, Text,\\
 &Dynamic, \neg Sec, Database, Active, DataTransfer)
\end{align*}

\section{\uppercase{Discussion}}
Table \ref{satisResults} shows the satisfaction results for the scenario presented in the previous section, where $d$ is the number of configuration choices made by the stakeholder and to which he/ she assigned the given importance degree shown in the row title, $r$ is the retained number among them in the final configuration, and $s$ gives the satisfaction rate for that importance degree computed by dividing $r$ by $d$. Each column is dedicated to as stakeholder. The last column gives the statistics for the final configuration. It shows that 4 configuration choices among 5 who were assigned 5 as importance degree were retained and 5 configuration choices among 7 with 4 as importance degree are present in the final configuration. Besides, no configuration choice with 1 as importance degree was retained, which makes sense.  The weighted satisfaction shows that that final configuration globally satisfied 72\% of the stakeholders configuration choices. When each stakeholder is taken separately, we find that $Stk5$ is fully satisfied (100\%), while $Stk2$ is not satisfied at all (0\%) even though he/she assigned the highest importance degrees to his/her configuration choices (5, 4 and 3). The scenario example was deliberately chosen to  show that the retained configuration choices depend not only on what the stakeholder assigns (the highest importance degree), but also on what the other stakeholders chose (i.e. the number of participating configuration choices in a conflict resolution), what importance degrees they assigned to their choices. After all, the IRatePL2C approach is based on the awareness of stakeholders about how conflicts are resolved.

\begin{table*} [htb!]
 \centering
        \caption{Satisfaction results for the scenario example.}  \centering
       \begin{tabular}{p{15cm}}
       \includegraphics [width=14cm]{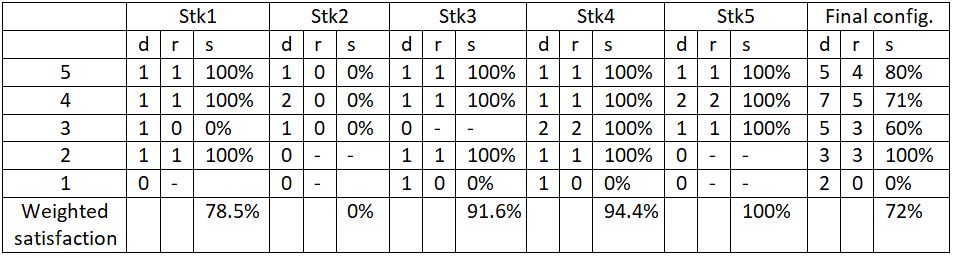}
\end{tabular}
\label{satisResults}
\end{table*} 

Regarding the complexity, the proposed solution is polynomial. Each of its sub-processes relies on walking through lists and constructing new lists. It depends on the number of involved stakeholders, the number of features of the product model, the number of related constraints and the number of features per constraint. 

Previous works that may be compared to ours, i.e. focusing on collaborative configuration of product lines and allowing a free-order process and taking into account stakeholders preferences in the conflicts resolution, have exponential complexity since they build a set of combinations based on a given set of features (e.g. SAT solvers, MCS computing); such problems are known to be NP-complete.

In \cite{stein2014}, stakeholders preferences regarding the conflict resolution are expressed through (1) hard constraints including positive \/ negative hard constraints used to indicate features that must be part \/ excluded of the configuration; and (2) soft constraints in the form of real number between -1 and 1 associated with (un)desired feature; 0 indicates indifference. The resolution process starts by checking between all stakeholders hard constraints, which may require a backtracking to review these hard constraints. Second, it generates the set of all valid configurations satisfying the hard constraints using SAT solver. Then, an optimal configuration is chosen after evaluating the obtained configurations with set of social choice strategies. 

As to \cite{Ochoa15}, stakeholders preferences are expressed by means of non-functional attributes specified according to a defined domain-specific language. A Constraint Satisfaction Problem (CSP) is generated using model-based transformation from the feature model, the stakeholders’configurations, and the defined preferences. The CSP is then solved to generate a set of valid configurations satisfying all constraints.

In \cite{viet2022}, stakeholders preferences are represented by means of Boolean numbers: 1 means want to include the feature and 0 means want to exclude the feature. The feature model is converted into CSP using model-based transformation. Stakeholders preferences are aggregated and the constraints solver finds a solution in case they are consistent. Otherwise, minimal conflict sets are determined using QuickXPlain algorithm. Minimal set of repair actions are then identified restore consistency in the individual stakeholder preferences and the optimal recommended one is based on the evaluation of a set of social choice strategies in the line of \cite{stein2014}.

Finally, stakeholders preferences in \cite{jss2023} are expressed through substitution rules (e.g. the most complete product, the simplest product) and used when one or more configuration decisions could not be retained in case of conflict. Stakeholders partial configuration (including desired and undesired features) are merged. The conflicts resolution strategy relies on the computation of the Minimal Correction Subsets based on Minimal Unsatisfied Subsets representing the list of detected conflicts. Substitution rules picked by stakeholders allow to identify the MCS(s) that resolve the identified conflicts.

\section{\uppercase{Conclusion}} \label{sec:conclusion}
In this paper, we presented a new approach to resolve conflicts in the context of collaborative configuration of product lines. Its process allows stakeholders to freely configure the product line model and takes into account their preferences to resolve conflicts expressed through an importance degree assigned to each explicit configuration choice. To reach its aim, IRatePL2C proceeds in steps; in each step, conflicts are detected, resolved and the intermediate configuration is updated accordingly before starting the following step. This allows to reduce the number of conflicts and reach easier to a valid solution. The approach does not prevent to have a completely dissatisfied stakeholder (i.e. all his/her configuration choices are not included in the final configuration), as the solution depends on the choices of all stakeholders regarding the features to include \/ exclude and the importance degrees they assign. Further empirical investigation is needed to reveal some ``tips" to avoid such situations. This is the subject of our future work.

\bibliographystyle{apalike}

\end{document}